\documentclass[pra,showpacs,floatfix,amsmath,twocolumn,amsfonts]{revtex4}

\def\ds{\displaystyle}

\usepackage{amsmath,amsthm}
\usepackage{epsfig}
\usepackage{amssymb}
\newcommand{\tom}{\tilde{\omega}}

\begin{document}

\title{Stationary localized modes in the quintic nonlinear Schr\"{o}dinger equation with a periodic potential}

\author{G. L. Alfimov$^1$}
%\email{galfimov@yahoo.com}
\author{V. V. Konotop$^{2}$}
%\email{konotop@cii.fc.ul.pt}
\author{P. Pacciani$^{2}$}
%\email{pacciani@cii.fc.ul.pt}
\affiliation{$^1$  Moscow Institute of Electronic Engineering,
    Zelenograd, Moscow, 124498, Russia
\\
$^2$Centro de F\'{\i}sica Te\'orica e Computacional,  Universidade de
Lisboa, Av. Prof. Gama Pinto 2, Lisboa 1649-003, Portugal and
Departamento de F\'{\i}sica, Faculdade de Ci\^encias,
Universidade de Lisboa, Campo Grande, Ed. C8, Piso 6, Lisboa
1749-016, Portugal.
}

\begin{abstract}

We consider the localized modes (bright solitons) described by one-dimensional quintic nonlinear Schr\"{o}dinger equation with a periodic  potential. In the case of attractive nonlinearity we deduce sufficient conditions for collapse. We show that there exist spatially localized modes with arbitrarily large number of particles. We study
such solutions in the semi-infinite gap (attractive case) and in the first gap (attractive and repulsive cases), and  show that a nonzero minimum value of the number of particles is necessary for a localized mode to be created. In the limit of large negative frequencies (attractive case) we observe ``quantization'' of the number of particles of the stationary modes. Such solutions can be interpreted as coupled ``Townes'' solitons and appear to be stable. The modes in the first gap have numbers of particles infinitely growing with frequencies approaching one of the gap edges, which is explained by the power decay of the modes.  Stability of the localized modes is discussed.

\end{abstract}

\pacs{03.75.Lm, 03.75.Kk, 03.75.-b}
\maketitle

\section{Introduction}

The quintic nonlinear Schr\"odinger (QNLS) equation
\begin{eqnarray}
\label{QNLS}
 iu_t+u_{xx}-V(x)u- \sigma |u|^{4}u =0\ ,
\end{eqnarray}
where either $\sigma=+ 1$ (the repulsive case) or  $\sigma=-1$ (the
attractive case), and $V(x)$ is an external potential, is a standard
model of the nonlinear physics.
This model can be viewed as the evolution equation $u_t=i\delta H [u]/\delta u^*$ generated by the Hamiltonian
\begin{eqnarray}
\label{energy}
    H[u]=\int_{-\infty}^{\infty}\left[|u_x|^2+V(x) |u|^2 +f(|u|^2)\right] dx
\end{eqnarray}
 where the nonlinearity
$f(|u|^2)$ has the second order term of the Taylor expansion equal to
zero, i.e. $f(|u|^2)=(\sigma/3)|u|^6+{\cal O} (|u|^8)$ (here we
took into account that $f(0)=0$, necessary for convergence of the integral, and dropped the unessential term $f'(0)|u|^2$ which results in adding a constant to the Hamiltonian). In
particular, this is the case of a slow envelope of a small amplitude
wave in a medium with weak dispersion and nonlinearity having specific
dependence on the wave intensity.

Recently, renewed interest in the model (\ref{QNLS}) was originated by the suggestion of its
use for description of a one-dimensional gas of impenetrable bosons~\cite{Kolom}.  
In the later studies~\cite{limitations}, limitations  of the
mean-field approach have been discussed. More specifically, it has been shown that the model fails at relatively small number of atoms and in description of the gas dynamics. Nevertheless, it has been rigorously proven in~\cite{LS}, that in the limit of large number of particles, there exist a domain where the stationary version of Eq. (\ref{QNLS}) [see (\ref{eq_phi}) below],
corresponding to the energy  functional given by the integral
(\ref{energy}) with $f(|u|^2)=(\pi^2/3) |u|^6$ (provided $V(x)$ has a
homogeneously growing part), does describe the ground state of the
  Tonks-Girardeau gas~\cite{TG}. 

Let us now assume, without loss of generality, that  $V(x)$ is a $\pi$-periodic function, $V(x)=V(x+\pi)$. This means that the energy, and in particular $V(x)$, is measured in units of the recoil energy $E_R=\hbar^2\pi^2/(2md^2)$, where $m$ is a mass of an atom and $d$ is the lattice constant, and time and spatial coordinate in the dimensionless Eq. (\ref{QNLS}) are measured    in the units $E_R/\hbar$ and $d/\pi$, respectively. Then, we observe that the number of real atoms  of the Tonks-Girardeau gas, which we denote as 
${\cal N}$, is connected with the integral of motion of the dimensionless model (\ref{QNLS}) $N=\int |u(x,t)|^2dx$ (hereafter
we drop limits in integrals over the real axis) by the formula ${\cal N}=N/\pi$ (for the sake of  
brevity below $N$ is also referred to as a number of particles).  

Next, we recall that in the case of the cubic nonlinear Schr\"{o}dinger equation, where
$f(|u|^2)\propto |u|^4$, with a periodic potential having amplitude of
the unity order, the quantity $N$ for a gap soliton is of
order of a few units~\cite{BK}, and that localized solutions of the
QNLS equation (\ref{QNLS}) reported in earlier publications~\cite{AS}
have also $N$ of order of one (see also the results presented below). The respective modes do not belong to the domain of the applicability of the model (\ref{QNLS}) in the theory of the Tonks-Girardeau gas (i.e. do not satisfy the necessary condition $N\gg 1$)~\footnote{The authors acknowledge a referee, who has drawing their attention to this issue.}. This rises  a natural question: can localized modes (or gap solitons) of the  model
(\ref{QNLS}) have any physical relevance to the theory of the Tonks-Girardeau gas?

In the present paper we will give a positive answer to this question, which
will be based on the argument, that the QNLS equation with a periodic
potential allows storage of an infinite number of particles, i. e.  allows lcalized mode solutions corresponding to the limit $N\gg 1$.

Meantime,  the model (\ref{QNLS}) appears also in  the quasi-one-dimensional limit of
the generalized Gross-Pitaevskii equation when two-body interactions
are negligibly small (what can be achieved, say by means of the
Feshbach resonance) and the chemical potential is approached by
$\mu\approx g_2n^2$ with the coefficient $g_2$ depending on the
three-body interactions and $n$ being the atomic density~\cite{AS,BKP} (see also~\cite{PS}). In that case, the total
physical number of particles is computed as ${\cal
N}=\sqrt{\frac{3\hbar^2a_\bot^4}{2\pi^2m |g_2|}}N $ (here $a_\bot$ is a
transverse linear oscillator  length, which must be  much bigger than
the scattering length $a_s$: $a_\bot\gg |a_s|$). Also in this case
$\sigma=$sign$\,(g_2)$.

The homogeneous case $V(x)=0$ of Eq. (\ref{QNLS}) has been
extensively studied (see e.g.\cite{collapse}). It is known that in the
attractive case Eq. (\ref{QNLS}) describes two specific scenarios of
the evolution. The first scenario corresponds to collapse when a smooth
solution ceases to exist and its amplitude infinitely grows in a finite
time. The second scenario is dispersion of a solution, when amplitude
of initially localized solution decays and the area of its localization
grows. These two regimes are "separated" by the soliton solution (see also Sec.~II below)
\begin{eqnarray}
\label{TS_V_0}
u(t,x)=\frac{3^{1/4}(-\omega)^{1/4}}{\sqrt{\cosh(2\sqrt{-\omega}x)}}\ e^{-i\omega t}\,.
\end{eqnarray}
Here $\omega$ is a dimensionless real constant parameterizing the solution.
As it is clear solution (\ref{TS_V_0}) exists for $\omega<0$. Bellow
$\omega$ will be referred to as frequency, while in the context of the background solution within the framework of the  mean-field theory of the quantum gases~\cite{TG,LS} $\omega$
determines the   chemical potential $\mu$, given by $\mu=E_R\omega/\hbar$.

In the repulsive case neither bright soliton solutions nor collapse regime can occur.

Presence of the periodic potential $V(x)$ in the model
greatly enriches its behavior.    In
Ref.~\cite{AS} it has been reported the existence of unstable solutions
which allow one to characterize the delocalizing transition~\footnote{In~\cite{AS} such solutions
were termed "gap-Townes" solitons. In the present paper we however use
different terminology, which emphasizes existence of various branches
of one-parametric families of solutions having different mathematical
properties (see the text for the definition).} and respectively the
existence of the minimal and maximal values of the number of particles
corresponding to existence of localized solutions beyond which the
solution is either collapsing or dispersing.

It turns out that the earlier results can be  significantly
generalized. In particular, in the present paper we show that in the
attractive case and for frequencies laying in a semi-infinite gap
solitary waves represent an {\em infinite number of branches of the
solutions} which can be parametrized by the number of particles. We
also show that the presence of the lattice significantly affects
properties of the model, on the one hand changing  sufficient
conditions for the collapse and on the other allowing one to store {\em
an arbitrary} number of particles either in a form of a soliton chain
or in a single soliton. We compute the bifurcation diagram for the
lowest branches of such solutions and describe a phenomenon of
"quantization" of the number of particles in the limit of large
negative frequencies in the attractive case. We also describe families
of the gap-soliton solutions showing great differences with the case of
the cubic nonlinear Schr\"odinger equation.

\section{Statement of the problem}

We assume that $V(x)$  is bounded, $|V(x)|\leq V_0$ ($V_0$ is a positive constant), $\pi$-periodic,  and even, $V(x)=V(-x)$, with a local minimum placed at $x=0$. We seek for stationary localized solutions in the form $u(x,t)=\exp(-i\omega t)\phi(x)$ where $\omega$ is a frequency. The function $\phi(x)$ can be considered real and satisfying the equation
\begin{eqnarray}
\label{eq_phi}
\phi_{xx}-\sigma\phi^5+\left[\omega-V(x)\right]\phi=0
\end{eqnarray}
with zero boundary conditions: $\lim_{|x|\to\infty}\phi(x)=0$. A localized solution decays at infinity, thus  reaching spatial domains where the nonlinearity becomes negligible. Therefore  its asymptotics are determined by the linear eigenvalue problem
\begin{eqnarray}
\label{eigen_probl1}
\frac{d^2 \varphi_{\alpha q}}{dx^2}+\left[\omega_{\alpha}(q)-V(x)\right]\varphi_{\alpha q}=0\,
\end{eqnarray}
which has a band spectrum~\cite{MW} $\omega_\alpha(q) \in \left[\omega_\alpha^{(-)},\omega_\alpha^{(+)}\right]$  with $\alpha=1,2,\dots$ referring to a band number and   ``+'' and ``--'' standing respectively for the upper and lower boundaries of the $\alpha$-th band.
$\varphi_{\alpha q}$ are Bloch functions and   $q$ is the wave number in the reduced Brillouin zone, $\ds q\in [-1,1]$.  An interval $\left(\omega_{\alpha}^{(+)},\omega_{\alpha+1}^{(-)}\right)$ represents the $\alpha$'s gap. The lowest gap, which we designate by $\alpha=0$, is semi-infinite, $\left(-\infty,\omega_1^{(-)}\right)$.  Since the asymptotics of localized solution of Eq.(\ref{eq_phi}) is determined by (\ref{eigen_probl1}), there exist no localized solutions of Eq.(\ref{eq_phi}) for $\omega$ lying in a band and only for $\omega$ lying in a gap the possible.

\begin{figure}[ht]
%\vspace{0.5 true cm}
\includegraphics[width=5cm]{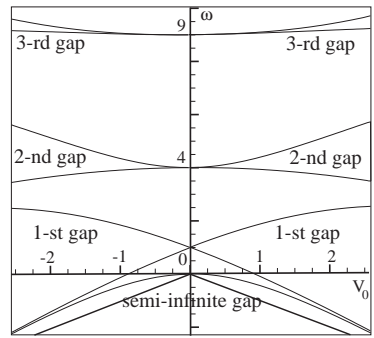}
\caption{The structure of the gaps for Eq.(\ref{eigen_probl1}),
$V(x)=-V_0\cos 2x$. The triangular area in the lower part of the figure
is the area $\omega<V_0$ where no localized
solutions exists in repulsive case (see the text).} 
\label{figzero}
\end{figure}

As we already mentioned, an important characteristics of the solution is the integral $N$, which in low-dimensional condensate applications is proportional to the number of particles of the condensate, therefore below we refer to $N$ as to a number of particles.

In the attractive case and in the absence of the periodic potential, $V(x)\equiv 0$, the number $N_*=\sqrt3 \pi/2$ separates the solutions with $N>N_*$ which may collapse  and solutions with $N<N_*$ for which no collapse can
occur~\cite{W1}.  Specifically, $N_*$ corresponds to the exact solution (\ref{TS_V_0}) and the frequency, $\omega<0$, can acquire an arbitrary negative value. This situation is similar to one discovered in the context of the nonlinear optics for the 2D cubic Schr\"{o}dinger equation having the stationary solution, separating collapsing and dispersing pulses. That solution was discovered in 1964 \cite{T} and today is referred to as Townes soliton. Using this analogy, the stationary solutions of (\ref{QNLS}) with $\omega<\omega_1^{(-)}$, i.e. in the semi-infinite gap, having the smallest number of particles at given $\omega$, will be referred to as {\em Townes solitons} (TSs). Thus, by TSs we call the one-parametric family of solutions corresponding to the lowest branch in the semi-infinite gap (see the discussion below and  Figs.~\ref{figone} and \ref{figtwo}). In the limit $V(x)\equiv 0$ TSs are transformed in the solitons given by (\ref{TS_V_0}). If $V(x)\neq 0$ stationary localized solutions of (\ref{QNLS}) with $\omega$ lying in a finite gap, $\omega\in \left(\omega_{\alpha}^{(+)},\omega_{\alpha+1}^{(-)}\right)$, may exist. Also, the new solutions, with the number of particles bigger than the number of particles of TSs, may exist in the semi-infinite gap. All such solution disappear in the limit $V(x)\equiv 0$, and therefore we will call them   {\em gap solitons}.

\section{Sufficient condition for collapse in the attractive case ($\sigma=-1)$}

From simple intuitive arguments it follows that the stability of an initially  strongly localized wave packet, (i.e. localized on distances much smaller than the lattice period), should not be significantly affected by the lattice. Indeed, boundness of  $\int V \phi^2dx$ and use of the known arguments for obtaining $N_{*}$ ~\cite{W1,collapse}  allow one to conclude  that the presence of a periodic potential does not affect the fact that {\it no collapse can occur if $N<N_{*}$}.

To deduce  a sufficient condition for the collapse for Eq.(\ref{QNLS}) we introduce two momenta,
\begin{eqnarray*}
&&y(t)=\int x^2|u(x,t)|^2 dx\, ,\\
&&z(t)=\mbox{Im}\int x u(x,t)\bar{u}_x(x,t) dx\,
\end{eqnarray*}
and the energy functional
\begin{eqnarray}
    E=\frac 12 \int\left(|u_x|^2-\frac 13 |u|^6+V |u|^2\right)dx\,,
\end{eqnarray}
which is a first integral of Eq.(\ref{QNLS}). Next, following the standard steps~\cite{collapse}, and assuming that the solution exists for $0\leq t<T$, we obtain
\begin{eqnarray}
 \label{rel_y}
\frac{d^2 y}{dt^2}=-4\frac{dz}{dt}
  = 16E-   4
    \int (2V+x V_x)|u|^2dx
\end{eqnarray}
Let us show that the conditions
\begin{eqnarray}
\label{ID}
    z(0)\geq 0\quad\mbox{and}\quad\eta=-4E_0-V_1y^{1/2}(0)N^{1/2} >0\ ,
\end{eqnarray}
where $E_0=E+V_0N/2$ and $V_1=\sup_x|V_x(x)|$, are {\em sufficient conditions} for collapse in presence of periodic potential.

First, one can show that $y(t)$ is a decreasing positive function, i.e. $0<y(t)<y(0)$.  Indeed, define $v(t)\equiv y(t)-y(0)$ and   obtain from
(\ref{rel_y})
\begin{eqnarray}
\label{auxil}
    \frac{dv(t)}{dt}\leq -4\eta t -4z(0)+\frac12\int_0^tv(s)ds\equiv F(t)\ .
\end{eqnarray}
Then $dv/dt<0$ for all $t\in[0,T)$. Indeed, let us assume  that there exists $t_*<T$ such that $dv(t_*)/dt=0$, while $dv/dt<0$ for $t<t_*$. From the definition of $F(t)$ it follows that $F(0)\leq0$, $dF/dt=-4\eta+v/2<0$ for $t\in[0,t_*)$ and, consequently, $F(t_*)<F(0)<0$. This contradicts the condition $F(t_*)\geq 0$ which
follows from (\ref{auxil}). Thus, $v(t)$ decreases at $t\in[0,T)$. In addition, Eq. (\ref{rel_y}) implies that
\begin{eqnarray}
\label{estim1}
    y(t)-y(0)
     \leq  (8E_0+2V_1y^{1/2}(0)N^{1/2}) t^2 -4z(0)t
\end{eqnarray}
Thus we have outlined the proof of the following statement:

Theorem -- {\it
 Let conditions   (\ref{ID}) be satisfied.
  Then,  $y(t)\leq y(0)-2\eta\,t^2$ and collapse
  occurs at finite time \mbox{$T\leq
\left(y(0)/2\eta\right)^{1/2}$}.
 }

Comparing the sufficient conditions (\ref{ID}) with the conditions for collapse for homogeneous case $V_1=0$ we conclude that in the presence of a periodic potential the collapse occurs at smaller values of energy $E<0$. This result has simple physical explanation. Indeed, for atomic distributions involving several local minima, the respective maxima act act as a repulsive force diminishing the effect of the inter-atomic attraction.

\section{A minimum number of particles for a localized solution}

When $N<N_*/\sqrt{3}$, multiplying (\ref{eq_phi}) by $\phi$, integrating, and  using the Gagliardo-Nirenberg inequality in the form $\int\phi^6 dx\leq(3N^2/N_*^2)\int\phi_x^2dx$ (see ~\cite{GN}) we obtain the estimate
\begin{eqnarray}
\label{ineq}
    \int\phi_x^2dx\leq(\omega +V_0)N\left(1+\frac{3}{2}\frac{N^2}{N_*^2}(\sigma-1)\right)^{-1}
\end{eqnarray}
This inequality has the following nontrivial consequence:
{\em For the stationary solutions of Eq.(\ref{QNLS}) the limit of a small number of particles, $N\to 0$, if available, implies smallness of the amplitude of the function, $\sup_x|\phi|\to 0$. The stationary solutions at $\omega<-V_0$ (if any) have $N>N_*/\sqrt3$.}

This statement follows immediately from (\ref{ineq}) and the inequality $(\sup_x\phi)^4\leq 4N \int\phi_x^2dx$. Thus, in order to establish the existence of a lower bound for the number of particles necessary for creation of a stationary localized solution, we address the small amplitude limit of Eq. (\ref{QNLS}).

The respective solutions bifurcate from the linear band, and thus can be described in terms of the effective mass approximation~\cite{BK}.
Thus we look for a solution of (\ref{QNLS}) in a form $\psi=\epsilon^{1/2}\psi_1+\epsilon^{3/2}\psi_2+\cdots$, where $\epsilon\ll 1$ and  $\psi_j$ are functions of slow variables $x_p=\epsilon^p x$ and $t_p =\epsilon^p t$. Then, following the standard procedure, we obtain that $\psi_1=A(x_1,t_2)\varphi_{\alpha}(x_0)\exp(-i\omega_{\alpha}t_0)$, where $\omega_{\alpha}$  stands either for $\omega_{\alpha}^{(+)}$ (if $\sigma=1$) or for $\omega_{\alpha+1}^{(-)}$ (if $\sigma=-1$) and $\varphi_{\alpha}$ stands for the respective Bloch function [see (\ref{eigen_probl1})]. The slow envelope $A(x_1,t_2)$ solves the QNLS
equation
\begin{eqnarray}
\label{TG}
    i\frac{\partial A}{\partial t_2}+
    \frac{1}{2M_\alpha}\frac{\partial^2 A}{\partial x_1^2}-\sigma \chi_\alpha |A|^4A=0
\end{eqnarray}
where $\chi_\alpha=\int_{0}^{\pi}|\varphi_\alpha (x)|^6dx$  is the effective nonlinearity and $M_\alpha^{-1}=d^2\omega_\alpha(q_0)/dq^2$ is  the effective mass ($q_0=0$ and $q_0=1$ for the center and the boundary of the Brillouin zone, respectively). The soliton solution of (\ref{TG}) is known:
\begin{eqnarray}
\label{A}
A=\frac{3^{1/4}\epsilon^{1/2}}{\chi_\alpha^{1/4}}
 \frac{\exp(-i\sigma\epsilon^2 t)}{\sqrt{\cosh\left(\epsilon x\sqrt{8|M_\alpha|}\right)}}\ .
\end{eqnarray}
Thus $\epsilon^2$ can be interpreted as  a frequency detuning to the gap, $\epsilon^2=|\omega-\omega_\alpha|$, and the number of particles
\begin{eqnarray}
\label{N}
N\approx \displaystyle{{\cal N}=\epsilon\int
|A|^2dx=\frac{\pi\sqrt{3}}{2\sqrt{2|\chi_\alpha M_\alpha|}}}\ ,
\end{eqnarray}
 does not depend on the detuning, but is determined {\em only} by the lattice parameters. This contrasts with the situation which takes place in the case of nonlinear Schr\"odinger equation  with cubic nonlinear term: the number of particles $N$ there tends to zero together with the detuning: $N\propto\sqrt{|\omega_\alpha-\omega|}$ (see e.g. \cite{BK}).

\begin{figure}[ht]
\vspace{1.3 true cm}
\includegraphics[width=\columnwidth]{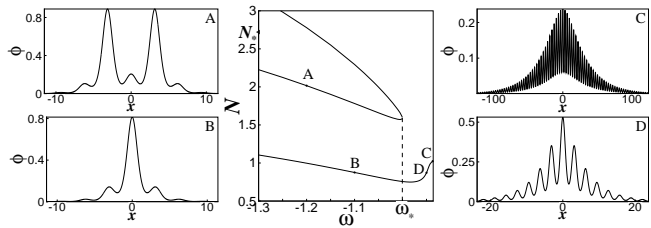}
\caption{Number of particles $N$ vs frequency  $\omega$ for even TSs at
$V_0=3$ (the central box). In the side boxes explicit forms of
solutions are shown. The box letters correspond to the points on the
branches: (A) for $\omega=-1.2$, (B) for $\omega=-1.1$, (C) for
$\omega=-0.9374$ and (D) for $\omega=-0.95$. $\omega_*\approx 1.0005$
is the bifurcation point of the second and third lowest branches.
} \label{figone}
\end{figure}

 The above arguments prove the existence of the minimal number of particles necessary for the creation of stationary localized mode described by Eq.(\ref{QNLS}). Indeed, if $N\to0$ would be possible, it necessarily  should be reached at solutions (\ref{A}). The latter, however, does not allow such a limit. At the same time,
according to the theory presented, the obtained estimate for $N$ itself needs not be the smallest number of particles of the stationary solutions. And it is not, as is seen from examples of numerical studies shown in Figs.~\ref{figone},~\ref{figthree}, and ~\ref{figfour} where, like in all other simulations, we use the potential
\begin{eqnarray}
\label{V0}
V(x)=-V_0\cos(2x)
\end{eqnarray}
 and show only the three lowest branches.

\section{Semi-infinite gap, Townes solitons.}

\subsection{The attractive case.}

Fig.~\ref{figone} shows the behavior  $N$ {\it vs} $\omega$ in the
semi-infinite gap in the vicinity of its upper bound
$\omega_1^{(-)}\approx -0.937$. The number of particles corresponding
to TSs (the lowest branch) achieves its {\em non-zero minimum}, showing
nonmonotonic behavior. At large frequency detunings toward the gap new
branches of the solution appear. Two of them bifurcate at some point
$\omega=\omega_*$. Fig.~\ref{figone} also shows that increase
(decrease) of the number of particles in a TS necessarily lead neither
to collapse nor to spreading out of the solution. Indeed, simultaneous
change of the frequency and the number of particles, corresponding to
motion along the respective branch, simply modifies the stationary
solution, what is expressed in the reshaping of the mode: when
frequency approaches the gap a mode broadens and its amplitude
decreases (c.f. boxes B, C and D).

Existence of the local  minimum on the lowest curve  $N (\omega)$, at a
frequency shifted from the edge  toward the gap, does not contradict
the small amplitude approximation which gives (\ref{N}), thus
predicting $N$ to be constant. Indeed, the small parameter
$\epsilon^{1/2}\sim 0.1$ corresponds to the frequency detuning
$|\omega-\omega_\alpha|\sim 10^{-4}$ i.e. is extremely small, therefore
the region of validity of asymptotic expansions is also very narrow.
This is illustrated by the panels B, C and D in Fig.~\ref{figone},
showing that reshaping of a soliton (from the envelope soliton in C to
an intrinsic localized mode in B) occurs already at quite small
frequency detuning.  It is worth it to compare this result with the one
known for the cubic nonlinear Schr\"odinger equation, where the number
of particles is a monotonic function of the frequency
detuning~\cite{BK}: $A_\alpha\propto|\omega-\omega_\alpha|^{1/2}$ and
the small parameter is $\varepsilon$ instead of $\varepsilon^{1/2}$
obtained above.

As it is illustrated by the panels B,C, and D, the lowest branch describes a family of one-soliton solutions. In a similar way the upper branches can be associated with multi-soliton solutions. As an example in panel A of Fig.~\ref{figone} we show a two-soliton mode corresponding to the second lowest branch. In the point $\omega=\omega_*$ shown in the central panel, this mode bifurcates with a three-soliton solution (not shown here) corresponding to the next upper branch. In the same manner one can describe the whole set of infinite number of upper branches.

\begin{figure}[th]
\includegraphics[width=\columnwidth]{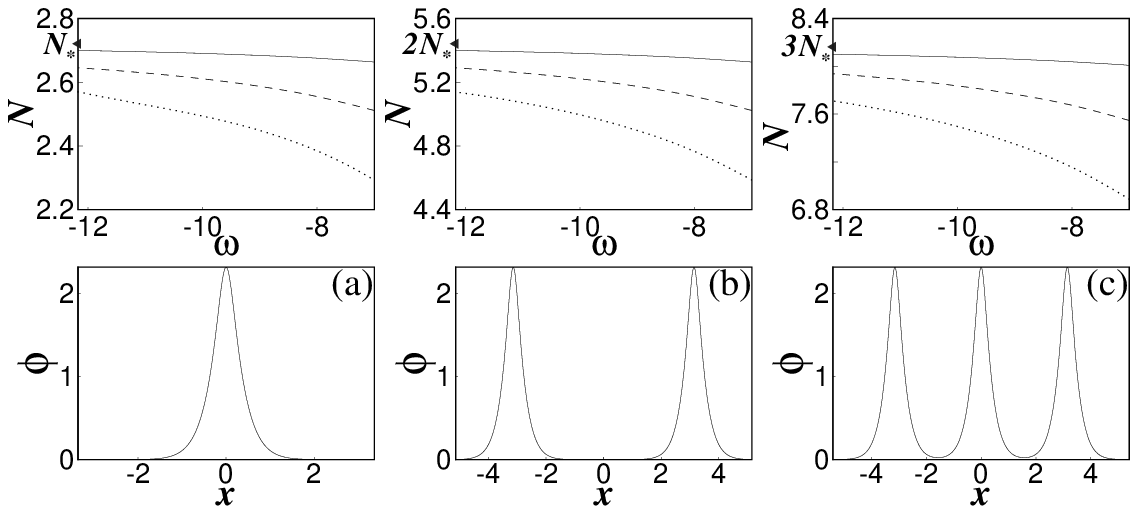}
\caption{Number of particles {\it vs} frequency for even TSs, at $V_0=1$ (solid line), $V_0=3$ (dashed line), and $V_0=5$ (dotted line), corresponding to first, second and third branches shown in Fig~\ref{figone}. In the boxes (a), (b), and (c) we  show explicit shapes  of the solutions for $V_0=3$ and $\omega=-12.175$.}
\vspace{1.5 true cm}
\label{figtwo}
\end{figure}

Although near the bifurcation point, the shapes of the modes collapsing with each other do not allow clear identification of $n$-soliton solutions, the introduced interpretation is well supported by the analysis of the modes at large negative frequencies, illustrated in Fig.~\ref{figtwo}. Indeed, independently  on the potential depth, solutions of all branches tend to chains of equidistant Townes solitons, the number of atoms of each one approaching $N_*$. While the lowest branch solution has its counterpart for the homogeneous QNLS equation, the coupled solitonic states of the higher branches do not exist at $V(x)\equiv 0$. This means that
a lattice of arbitrary depth can be treated as a singular perturbation allowing soliton binding. Our numerical studies allow us to formulate the following conjecture about "quantization" of the number of particles:

Conjecture 1 -- {\it  In the limit $\omega\to-\infty$ there exists an infinite number of different stationary solutions of the QNLS equation with a periodic potential, each of them having $nN_*$, where $n$ is an integer, number of particles.}

Recalling that the period of the potential is $\pi$ one observes form Fig.~\ref{figtwo} that the distances between coupled Townes solitons are integers of the lattice constant: it is $2\pi$  and $\pi$ in panels (b) and (c), respectively. One more property of the solutions one can observe in the upper panels of Fig.~\ref{figtwo}: approach of the solutions to a set of Townes solitons, as  $\omega\to -\infty$, is much faster for smaller amplitudes of the potential. This is explained by the fact, that the respective limit corresponds to the situation where the periodic potential can be viewed as a (singular) perturbation, and is achieved more rapidly for smaller amplitudes of the lattice.

Summarizing the results illustrated in Figs.~\ref{figone} and \ref{figtwo} we concludes that there exists no upper bound on the number of particles which can be loaded in lattice.

Finally, we notice  that the  whole lower branch in Fig.~\ref{figone} lies below the critical value $N_*$. This means that, unlike  the homogeneous QNLS equation~\cite{W1}, small enough perturbations of stationary solutions in presence of a periodic lattice do not result in collapsing behavior.

We checked the stability of the obtained solutions by performing direct numerical simulations of their dynamics governed by Eq. (\ref{QNLS}), after introducing a perturbation of the initial form $(1+\delta)\phi(x)$, where $\delta$ was $\pm 0.01$ and $\phi(x)$ was given by  one of the distributions shown in Fig.~\ref{figone} (A-C).
Performing computations until $t=1000$ we did not observe any significant change in the behavior of the modes~\footnote{After submission of the manuscript, long-time numerical studies of the mode (a), until $t=2000$ were performed by M. Salerno (personal communication). Those studies have shown that the mode  (a) is also unstable, what allows one to conjecture that the whole negative-slope part of the lowest branch is unstable, as in the standard situation described by Vakhitov-Kolokolov crietrion.}. In the case (D) , however a decrease of the amplitude ($\delta=-0.01$) resulted in spreading out of the solution, confirming the results reported earlier in~\cite{AS}. Increase of the amplitude, achieved by taking $\delta=0.01$, resulted in a oscillation of the shape of the mode, which was neither collapsing nor reaching any stationary state.

\subsection{The repulsive case.}

In the repulsive case one can easily prove that no localized solution of Eq.(\ref{QNLS}) can exist for $\omega<-V_0$. Indeed, from the Pohozaev identities for (\ref{eq_phi}) with $\sigma=1$, which read
\begin{eqnarray*}
&&  \int\left(\phi_x^2-\frac 13 \phi^6+\omega\phi^2-[V(x)+xV_x(x)]\phi^2\right)dx=0
    \\
&&  \int\left(\phi_x^2 + \phi^6- \omega\phi^2+V(x)\phi^2\right)dx=0,
\end{eqnarray*}
we deduce
\begin{eqnarray*}
    \omega N&=&\int\left(\phi_x^2+ \phi^6 +V(x)\phi^2\right)dx
    \\
    &\geq& \int V(x)\phi^2dx\geq -V_0 N
\end{eqnarray*}

In the case (\ref{V0}) the region where $\omega<-V_0$ does not cover the semi-infinite gap completely (see Fig.\ref{figzero}). In the non-covered part of the semi-infinite gap (a finite interval of $\omega$ existing for any fixed $V_0$) we performed a search of localized solutions using a code based on a specific version of the shooting method.  The obtained results allows us to formulate the following:

Conjecture 2 -- {\it No localized solution of Eq.(\ref{QNLS}) can exist in semi-infinite gap in the repulsive case.}

However we failed to prove the Conjecture 2 rigorously.

\section{Gap solitons of the first gap.}

The behavior of $N$ {\it vs} $\omega$ lying in the first gap is shown in Fig.\ref{figthree} (attractive case, odd modes) and Fig.{\ref{figfour} (repulsive case, even modes). As for the semi-infinite gap, one observes nonzero minima for the number of particles, which are necessary for creation of gap solitons and achieved at nonzero detuning toward the gap.  Another interesting feature that one can see on the both figures Fig.\ref{figthree} and
\ref{figfour} is that the numbers of particles for the localized modes {\it grow infinitely} when frequency approaches one of the gap edges.
This phenomenon is explained by the {\em algebraic} asymptotic of the gap soliton, which corresponds to the border of the band: $\phi\propto 1/x^{1/2}$ as $x\to \infty$. It is worth it pointing out here that, due to the sign of the effective mass, which in terms of the Eq. (\ref{TG}) would mean $\sigma M_\alpha\chi_\alpha>0$, it does not allow existence of small-amplitude bright solitons of the type (\ref{A}) in vicinity of lower bound of the first gap in attractive case and upper bound of the first gap in repulsive case, $V_0>0$. The following self-consistent recurrence procedure allows one to compute the basic term of algebraic asymptotics for the gap soliton corresponding to the border of 1-st gap as well as higher corrections to it.

Let us denote by $\tom_\alpha$ the respective gap boundary, such that $\tom_\alpha=\omega_{\alpha+1}^{(-)}$ if $\sigma=1$ and $\tom_\alpha=\omega_{\alpha}^{(+)}$ if $\sigma=-1$. Let $\varphi_0$ be the periodic Bloch function corresponding to $\tom_\alpha$ (notice that, compared with Sec. IV, here, for the sake of convenience, we change the notations of Bloch functions bordering a gap), $\varphi_0(x)=\varphi_0(x+2\pi)$,  normalized by the relation
$\int_{0}^{2\pi}\varphi_0^2dx=1$.  We seek a solution of Eq. (\ref{eq_phi}) in the form of the formal asymptotic series $ \phi=\gamma\left(\varphi_0/x^{1/2}+{\varphi_1}/{x^{3/2}} \cdots \right)\,,$ where $\gamma$ is a constant to be found. This leads to the recurrent formula for $\varphi_n$ ($n=1,2...$), the  first step of which leads to
\begin{eqnarray}
\frac{d^2\varphi_1}{dx^2}+\left[\tom_\alpha -
V(x)\right]\varphi_1=\frac{d\varphi_{0}}{dx}.
\end{eqnarray}

A solution of this equation  can be represented in the form
$\varphi_1=\tilde{\varphi}_1(x)+c_1\varphi_0(x)$, where $c_1$ is a constant and $\tilde{\varphi}_1$ satisfies the orthogonality condition $\langle\tilde{\varphi}_1,\varphi_0\rangle=0$. Considering the terms of the third order ($n=3$), i.e. those proportional to $x^{-5/2}$, we obtain the equation
\begin{eqnarray}
\frac{d^2\varphi_2}{dx^2}+\left[\tom_\alpha -
V(x)\right]\varphi_2=-\sigma\gamma^4\varphi_0^5+3\frac{d\varphi_1}{dx}-\frac 34\varphi_0
\end{eqnarray}
for which the orthogonality condition yields
\begin{eqnarray*}
    \sigma\gamma^4\int_{0}^{2\pi}\varphi_0^6~dx+\frac 34 \int_{0}^{2\pi}\varphi_0^2~dx-
    3\int_0^{2\pi}\varphi_0\frac{d\varphi_1}{dx}~dx=0
\end{eqnarray*}
Taking into account the representation of $\varphi(x)$ as well as normalization condition for $\varphi_0$ we compute
\begin{eqnarray}
    \gamma= \left(\frac{12\sigma\int_0^{2\pi}
    \varphi_0\frac{d\tilde{\varphi_{1}}}{dx}dx-3\sigma}{4\int_0^{2\pi}\varphi_0^6 dx} \right)^{1/4}.
\end{eqnarray}
Next, representing  $\varphi_2=\tilde{\varphi}_2+c_2\varphi_0$, and considering the terms corresponding to $x^{-7/2}$ we obtain
\begin{eqnarray}
\frac{d^2\varphi_3}{dx^2}+\left[\tom_\alpha -
V(x)\right]\varphi_3=5\frac{d\varphi_2}{dx}-5\gamma^4\varphi_0^4\varphi_1-\frac{15}{4}\varphi_1
\end{eqnarray}
Now, from the orthogonality condition  we obtain the coefficient $c_1$ (which is expressed by a bulky formula not presented here).

Continuing the described procedure yields the terms of the expansion for $\phi$ up to any order and all of them are unambiguously defined. Thus, the {\em logarithmic} divergence of the number of particles $N$ takes place as $\omega\to\tom_\alpha$,  which conforms to Figs.~\ref{figthree},~\ref{figfour}.
Here we notice that the similar expansion for the cubic NLS equation gives  the decay $\propto x^{-1}$, which implies finiteness of the number of particles for the solution on the gap boundary.
\begin{figure}[ht]
%\vspace{0.3 true cm}
\includegraphics[width=\columnwidth]{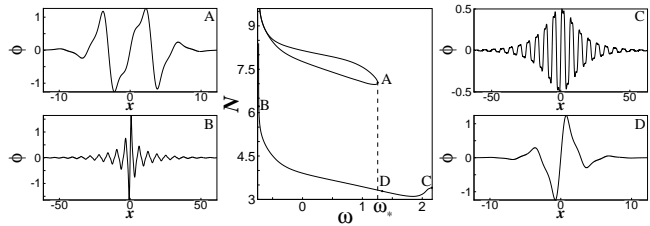}
\caption{ Number of particles  of odd modes {\it vs} frequency in the first gap, for $V_0=3$ in attractive case (central box). Mode profiles  in the points A ($\omega\approx 1.26$), B ($\omega=-0.73$), C ($\omega=2.15$), and D ($\omega=1.33$) are shown in the side  boxes. $\omega_*\approx 1.256$ is the bifurcation point of the second and third branches.}
\vspace{1.5 true cm}
\label{figthree}
\end{figure}

\begin{figure}[ht]
\vspace{0.3 true cm}
\includegraphics[width=\columnwidth]{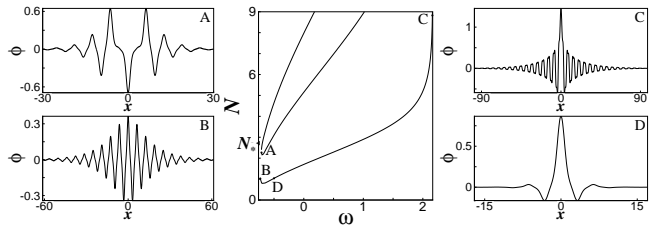}
\caption{ Number of particles  of even modes {\it vs}  frequency in the first gap, for $V_0=3$ in repulsive case. Mode profiles  in the points A ($\omega=-0.7105$), B ($\omega=-0.73$), C ($\omega=2.16$), and D ($\omega=-0.5$) are shown in the side  boxes. }
%\vspace{0.3 true cm}
\label{figfour}
\end{figure}

Returning to Fig.~\ref{figthree}, we  observe that all branches of the spectrum are located above the critical value $N_*$, which suggests the possibility of observing instability and collapse of the respective solutions. Fig.~\ref{figfive}a shows an example of such dynamics obtained by direct numerical simulations.
\begin{figure}[th]
\includegraphics[width=4.25cm]{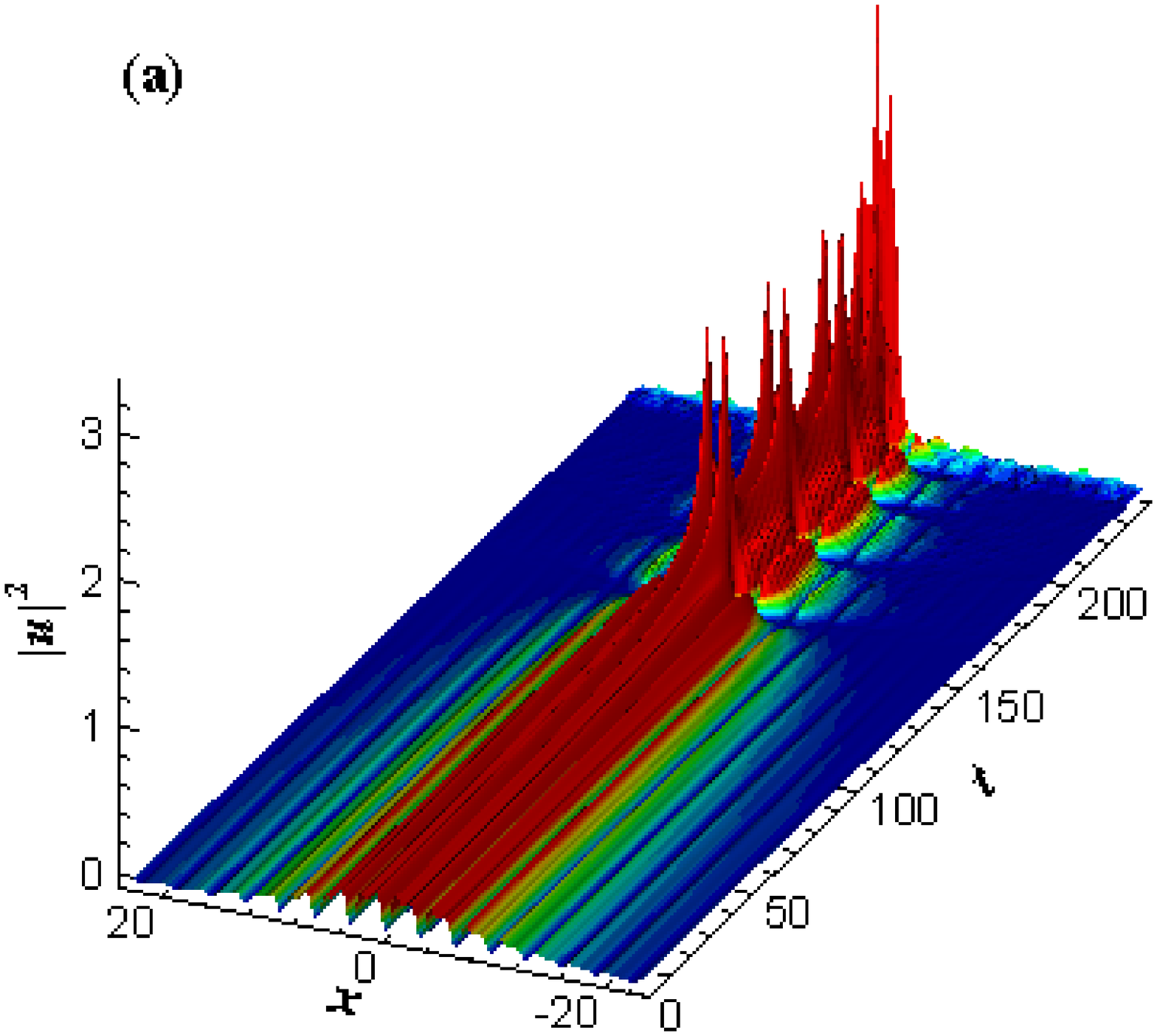}
\includegraphics[width=4.25cm]{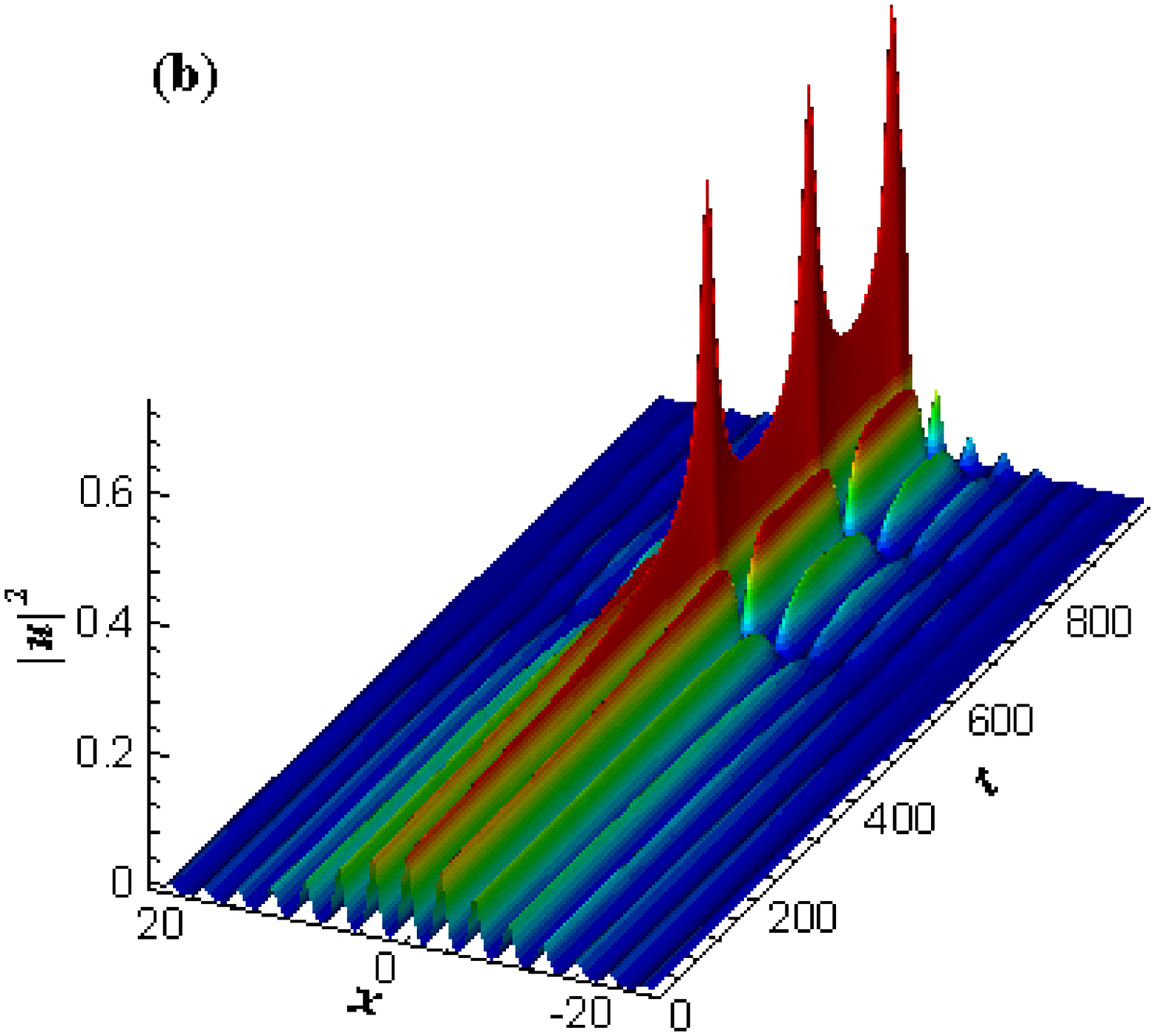}
\caption{Dynamics of gap solitons  in attractive (a) and repulsive (b) cases, starting with perturbed initial conditions given in Fig.~\ref{figthree}C and in Fig.~\ref{figfour}B multiplied by a factor $(1+\delta)$ with $\delta=0.01$. Negative $\delta$ in the both cases resulted in spreading out solutions (not shown). The simulation were carried out on the spatial interval (-60,60).}
\label{figfive}
\end{figure}

In both panels presented in Fig.~\ref{figfive},   the dynamics shows
three different stages of evolution, the first two being similar for
both attractive and repulsive cases. At the {\it first} stage, one
observes growth of the amplitudes and shrinking of the width of the
solutions. This "quasi-collapse" behavior is explained by the fact that
in both cases the dynamics is approximately described by the equation
(\ref{TG}), which is also an effective QNLS equation, and thus
corresponds to collapsing solutions. This is not an authentic collapse,
because after the particles are gathered mainly in a few (actually two)
potential wells, (\ref{TG}) is not applied any more and (roughly) the
system is similar to a dimer model. Oscillatory regime in such a model  
can be proven~\cite{Sacchetti} to originate the dynamics  
insensitive to the sign of the nonlinearity. This is the reason we
observe qualitatively similar oscillatory behavior of both systems
during the {\it second} stage of evolution. Finally, after several
(three in Fig.~\ref{figfive} a) oscillations in the attractive case we
observed collapse (the {\it third} stage). In the repulsive case,
oscillatory behavior was observed for longer times, and the last
(third) stage  corresponds to decay of the peak amplitude and
spreading out of the pulse.

\section{Conclusion}

In the present paper we have  shown that  a periodic potential results
in qualitative changes in  behavior of the quintic nonlinear
Schr\"odinger equation. It modifies the sufficient condition for the
collapse, requiring negative energies with larger absolute values,
allows the existence of infinite families of the localized solutions,
and can bind Townes solitons (the latter expressed by the conjecture
about quantization of the number of particles in the limit of large
negative frequencies). Either by  coupling Townes soliton or by
allowing existence of gap solitons, the periodic potential allows
storage of an infinite number of atoms. It also enforces the stability
of Townes solitons allowing them to have under-critical number of
particles.

The localized modes were shown to be very different from their counterparts in the cubic nonlinear Schr\"odinger equation, where the number of particles can go to zero and is bounded from above. Localized modes in the quintic nonlinear Schr\"odinger equation on the one hand require a minimal nonzero number of particles and on the other hand may have arbitrarily large number of atoms.

The  results reported in the present work, however, do not exhaust all stationary solutions. Speaking about solutions with of a definite parity we have restricted consideration only to the families originated by the lowest branch (i.e. by the one having the smallest number of particles). Possibility of existence of branches having different symmetry (say, odd solutions in the semi-infinite gap and even solutions in the first gap, in the case of attractive interactions) is left to be open question.

\acknowledgments

We are pleased to thank M. Salerno for a careful reading of the
manuscript and for a  number of useful comments about stability of the
modes, F. Kh. Abdullaev and  J. Brand for fruitful discussions, and V.
A. Brazhnyi for help with numerics. The work has been supported by the
grant POCI/FIS/56237/2004. PP was supported also by the FCT grant
SFRH/BD/16562/2004 and GA  thanks the President Program in Support for
Leading Scientific Schools in Russia (Project 4122.2006.2.)

\end{document}